\documentclass[runningheads]{llncs}

\usepackage{amsmath}
\usepackage{booktabs}
\usepackage{caption}
\usepackage{subcaption}
\usepackage{graphicx}
\usepackage[all]{nowidow}
\usepackage[utf8]{inputenc}
\usepackage{hyperref}
\usepackage[inline]{enumitem}
\usepackage{float}
\usepackage{longtable}
\usepackage{multicol}

\hypersetup{
    colorlinks=true,
    linkcolor=blue,
    citecolor=blue,
    urlcolor=blue
}

\begin{document}

\title{Proof of Concept as a First-Class Architectural Decision Instrument}

\author{Bruno Fernando Antognolli, Fabio Petrillo}

\institute{École de technologie supérieure (ÉTS), Montréal, Québec, Canada \\
\email{\{bruno-fernando.antognolli,fabio.petrillo\}@etsmtl.ca}}

\authorrunning{Antognolli and Petrillo}

\maketitle

\begin{abstract}
Proofs of Concept (PoCs) are widely adopted practices in software engineering. Despite their relevance, PoCs remain conceptually underdefined and methodologically ad hoc in both research and industry, with definitions and implementation approaches that often lack clarity and consistency.
This paper investigates the concept of PoCs with two complementary goals:\textbf{ (1)~to provide a refined definition and astructured framework} for PoC development grounded in a systematic review of
academic and grey literature; and (2)~\textbf{to position PoCs as \emph{first-class
architectural decision instruments}} rather than informal experiments or disposable artifacts.  Through a systematic review of academic and grey literature we identify the key characteristics, processes, associated with PoCs and expose a significant gap---the academic
literature describes PoC \emph{outcomes} but rarely its \emph{process}.  By synthesizing insights from diverse sources we propose a refined definition and a
lightweight, three-phase framework (planning, execution, decision-making) that
encompasses technical validation and explicit decision traceability. We also introduce the \emph{Undocumented Architectural Experiment} anti-pattern, arising when PoCs influence high-impact architectural decisions without leaving durable architectural knowledge.  We argue that elevating PoCs to first-class status improves decision quality, enhances traceability, and supports more systematic learning in architectural practice.
\end{abstract}


\section{Introduction}
\label{sec:introduction}

Software architecture is increasingly shaped by uncertainty and complexity
\cite{Esfahani2012,Sobhy-2021}.  Architects are routinely required to make
early, high-impact decisions in contexts characterized by incomplete
information, rapidly evolving technologies, and stringent non-functional
constraints---a situation particularly pronounced in cloud-native systems,
DevOps-oriented organizations, AI and machine learning pipelines, and
highly regulated domains such as finance and healthcare
\cite{Ameller2012,lakkarasu2025designing}.

In practice, Proofs of Concept (PoCs) are frequently employed to cope with this uncertainty.  Architects and development teams build PoCs to explore the
feasibility of technologies, to assess architectural strategies under realistic constraints, and to reduce risk before committing to large-scale design or implementation decisions \cite{baptista2024software}.  Typical questions addressed through PoCs include whether a given technology can meet performance or scalability requirements, whether an architectural approach integrates effectively with an existing ecosystem, or whether regulatory or operational constraints can be satisfied.

Despite their widespread use, PoCs remain poorly conceptualized in software
engineering research.  Dictionary definitions capture partial meanings---Collins
defines a PoC as \emph{``the stage during the development of a product when
it is established that the product functions as intended''} \cite{Collins2024},
while Oxford describes it as \emph{``evidence that shows that a business
proposal, design idea, etc., will work, usually based on an experiment or a
pilot project''} \cite{Oxford2024}---but neither captures the real meaning of
a software proof of concept.  The term is often used interchangeably with
prototypes, pilots, or minimum viable products, leading to confusion about
intent, scope, and expected outcomes \cite{larman2002applying,mistrik2012aligning,van2013learning}.
As a result, PoCs are frequently executed in an informal and opportunistic
manner, without explicit success criteria, architectural rationale, or
systematic linkage to the decisions they are meant to support.

This paper makes three complementary contributions.

\begin{enumerate}
\item \textbf{Definition and framework.}  We provide a refined definition of
PoCs in software engineering and a structured, lightweight three-phase framework
grounded in a systematic review of 20 academic and grey-literature sources.
\item \textbf{Architectural positioning.}  We argue that PoCs should be treated
as \emph{first-class architectural decision instruments}: short-lived,
risk-driven experiments whose primary outcome is evidence that supports or
refutes an architectural choice, not executable production code.
\item \textbf{Reusable artefact.}  We provide a publicly available, structured
PoC document template implementing the framework, enabling practitioners to
capture, structure, and trace the evidence underpinning architectural decisions.
\end{enumerate}

These contributions are organized around three research questions:

\begin{description}
\item[RQ1] \textit{What is a rigorous definition of a Proof of Concept in
software engineering?}
\item[RQ2] \textit{What structure and process characteristics make a PoC
architecturally effective?}
\item[RQ3] \textit{To what extent does the proposed framework improve PoC
planning, traceability, and decision quality compared to unstructured practice?}
\end{description}

The remainder of the paper is organized as follows.
Section~\ref{sec:background} situates PoCs in prior work.
Section~\ref{sec:methodology} describes the research methodology.
Section~\ref{sec:definition} presents the refined definition and addresses~RQ1.
Section~\ref{sec:characteristics} identifies key PoC characteristics and addresses~RQ2.
Section~\ref{sec:framework} introduces the structured framework and its document
template (Section~\ref{sec:framework-structure}).
Section~\ref{sec:casestudies} evaluates the framework through two case studies, addressing~RQ3.
Section~\ref{sec:architectural} develops the architectural perspective and
introduces the anti-pattern.
Section~\ref{sec:discussion} analyzes implications and limitations.
Section~\ref{sec:conclusion} concludes.\footnote{A position paper sharing the conceptual foundations of this work was presented at ICSE-Companion~2026~\cite{petrillo2026poc}.}

\section{Background and Related Work}
\label{sec:background}

\subsection{The Conceptual Ambiguity of PoCs}

One of the central problems surrounding PoCs in software architecture is their
conceptual ambiguity.  In both academic publications and industrial discourse,
the term ``PoC'' is used to denote a wide range of artifacts and activities.
In some cases a PoC refers to a quick technical experiment designed to test a
single hypothesis.  In others, it denotes a throwaway prototype, a reduced
implementation of a future system, or even a pilot deployed in a
production-like environment.

These interpretations differ substantially in scope, cost, lifespan, and intent,
yet they are often conflated under the same label.  This ambiguity leads to
misaligned expectations among stakeholders: architects, developers, managers,
and decision-makers may attribute different meanings and goals to the same PoC.

Some authors use the terms PoC, Prototype, and MVP synonymously
\cite{larman2002applying,fairbanks2010just,mistrik2012aligning,van2013learning,anderson2023design}. Others argue that these terms are distinct, highlighting differences in scope and purpose \cite{ingeno2018software,shrivastava2022solutions,appinventiv_2024,techmagic_2024}.
A key distinction lies in their \emph{primary purpose}: a PoC aims to
demonstrate the feasibility of a concept or technology, often with limited
features \cite{anderson2023design}, while a prototype may have broader goals
such as validating design ideas or guiding software implementation
\cite{mistrik2012aligning}.  Moreover, PoCs are typically short-lived and not
intended for final product use \cite{ford2017building,burns2018designing,hinchey2018software}, while prototypes may evolve into functional components.

\subsection{The Disposable Nature of PoCs}

The disposable nature of PoCs is a recurring theme in the literature.  They are often regarded as quick experiments designed to assess technical feasibility \cite{burns2018designing}.  This concept aligns with Martin Fowler's notion of ``sacrificial architecture'' \cite{fowler2015}, wherein PoC code is considered disposable due to design compromises made for rapid development \cite{ford2017building}.  While some argue that a PoC should never evolve into a deliverable due to the potential technical debt it accumulates \cite{beningo2022embedded}, others maintain that its purpose is solely to assess feasibility \cite{ford2017building,burns2018designing,hinchey2018software}.

This paper refines this view by distinguishing between two separable artifacts:
the \emph{implementation artifact} (source code, scripts, configurations) and
the \emph{experimental artifact} (the reproducible experiment itself, together with its scenarios, recorded metrics, and conclusions).  Only the implementation artifact needs to be discarded from the production codebase; the experimental artifact must be preserved.  A well-structured PoC should be \emph{reproducible}---its scenarios and environment specifications documented in sufficient detail that the experiment can be re-executed to yield consistent results.  

The idempotent quality of PoC transforms the PoC from a one-off activity into durable technical evidence capable of independently validating the architectural decision it informed.

\subsection{PoCs and Architectural Practice}

In contrast, established architectural practices such as Architecture Decision
Records~(ADRs) \cite{fowler2026adr}, risk-driven design, and methods like ATAM emphasize explicit rationale, trade-offs, and structured decision-making.  ADRs are short documents
that capture a single architectural decision together with its context, trade-offs,
and rationale~\cite{fowler2026adr}; they play a central role in the Advice
Process---a decentralised model in which anyone can make an architectural
decision, provided they first seek input from those affected and from domain
experts~\cite{fowler2023scaling}.  PoCs, despite their
central role in early architectural exploration, remain largely absent from
these frameworks---a missed opportunity to integrate experimentation and evidence
into architectural reasoning \cite{baptista2024software,brisals2024serverless}.

The informal treatment of PoCs has concrete consequences for architecture
practice.  Architectural decisions are frequently justified by PoC outcomes
whose assumptions, metrics, and limitations are undocumented or poorly
articulated.  This undermines decision traceability and makes it difficult to
revisit or reassess decisions when system requirements or constraints evolve.

\section{Research Methodology}
\label{sec:methodology}

The research methodology is structured around three key phases.

\begin{enumerate}
\item \textbf{Historical review.}  We trace the origins and early usage of the
term PoC within software engineering.  The concept originated in fields such as
aerospace and biomedical engineering before being adopted by the software
industry \cite{deckert1954analytic,welker1963seasat,mun1965polytechnic}.  The
first recorded mention of ``proof of concept software'' was in 1981, when NASA
used the term to validate software modules for the Integrated Programs for
Aerospace Vehicle Design (IPAD) \cite{bales1981structures}.

\item \textbf{Systematic literature review.}  We searched the Scopus database
for papers from top software engineering venues that explicitly address the PoC
\emph{process or methodology} (not just report PoC outcomes).  The query
retrieved 172 documents; after filtering for relevance and applying the
inclusion criteria in Table~\ref{tab-inclusionExclusionCriteria}, 0 academic
papers provided detailed PoC process descriptions---revealing a critical
research gap.

\item \textbf{Grey literature analysis.}  To fill this gap, we conducted a
systematic review of grey literature following the methodology proposed by
Garousi et al.~\cite{GAROUSI2019101}.  We analyzed 20 practitioner sources
(blog posts, white papers, technical reports) obtained through Google search
using the strings:
\texttt{"how to PoC" AND "software engineering"} and similar variants.
Stopping criteria were based on theoretical saturation and evidence exhaustion.
\end{enumerate}

\begin{table}[htbp]
\caption{Inclusion and exclusion criteria for academic literature}
\label{tab-inclusionExclusionCriteria}
\centering
\begin{tabular}{|p{5.5cm}|p{5.5cm}|}
\hline
\textbf{Inclusion criteria} & \textbf{Exclusion criteria} \\ \hline
Study is in the software engineering domain & Papers that focus solely on PoC
\emph{results} without describing the process \\ \hline
Study explicitly describes the PoC \emph{process} or methodology & Studies that
use ``proof of concept'' interchangeably with prototypes, pilots, or demos
without distinguishing the specific PoC process \\ \hline
\end{tabular}
\end{table}

\section{Defining PoC in Software Engineering}
\label{sec:definition}

The concept of PoC originated outside software engineering and was later adapted
without a clear, domain-specific definition.  This has caused misunderstandings
about what should be considered a PoC in software engineering, leading to
disagreements among different authors (Section~\ref{sec:background}).

An additional debate concerns whether the code developed in a PoC should be reused in future software versions. While developed in a PoC should be reused in future software versions. While authors advocate for code reuse, suggesting it can accelerate development \cite{van2013learning}, others argue that PoC code should be discarded, as its primary purpose is to test specific hypotheses within a limited context rather than laying the foundation for future development
\cite{ford2017building,burns2018designing,hinchey2018software,beningo2022embedded}.

Based on the evidence synthesized from academic and grey literature, we propose the following refined definition:

\begin{quote}
\textbf{A Proof of Concept (PoC) is a short-term
experiment that validates hypotheses within a limited scope by assessing
feasibility, mitigating risks, and supporting learning and decision-making.}
\end{quote}

This definition emphasizes several distinguishing properties: (i)~PoCs are
\emph{ephemeral}---they are time-constrained activities, not product increments;
(ii)~PoCs are \emph{hypothesis-driven}---each PoC tests concrete assumptions;
(iii)~PoC code is \emph{disposable}---it is not intended for production use;
and (iv)~PoCs are \emph{decision-oriented}---their primary output is evidence,
not software.

\begin{figure}[H]
    \centering
    \fbox{\includegraphics[width=10cm]{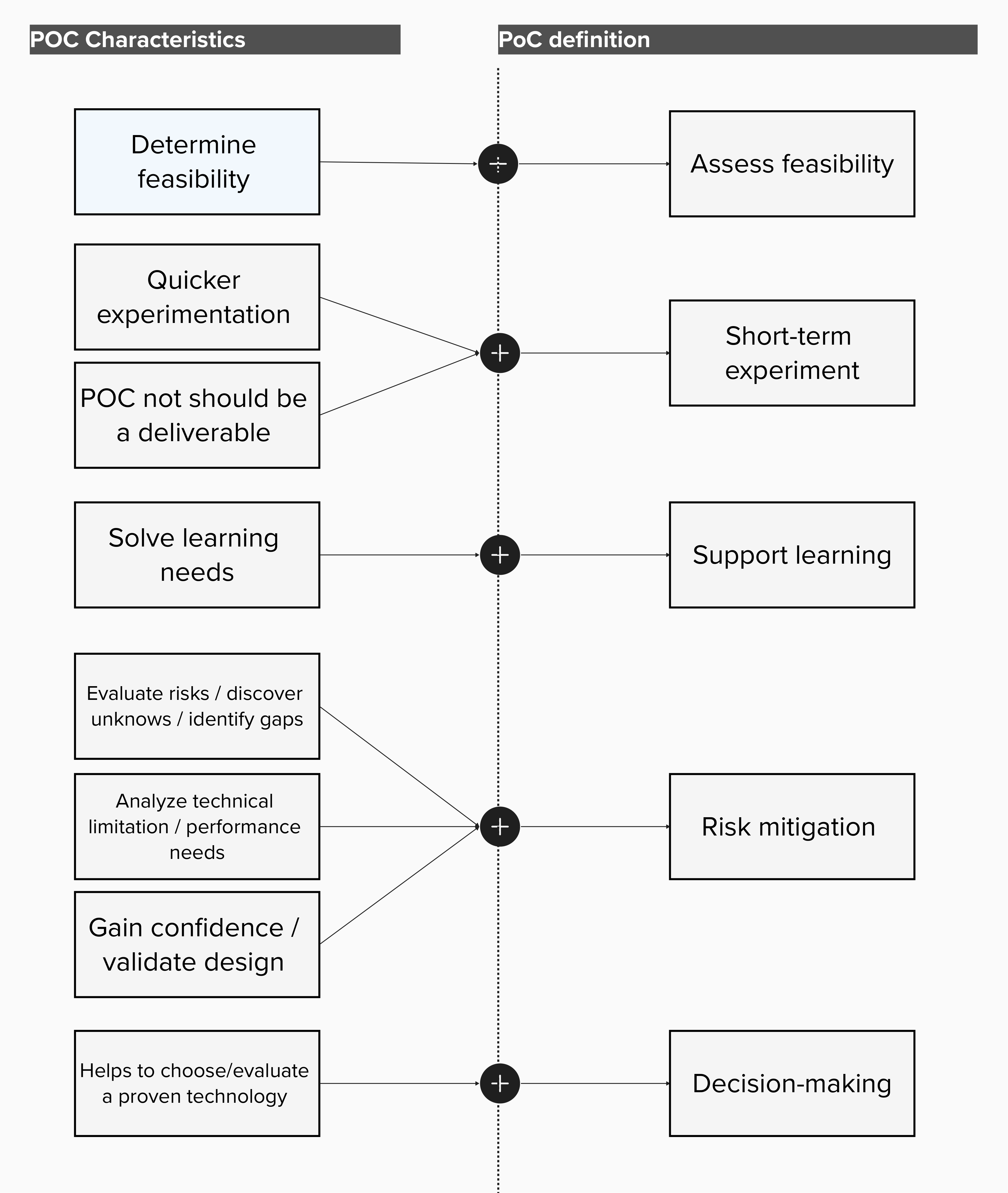}}
    \caption{PoC Characteristics and their link with a definition}
    \label{fig:pocCharacteristicsDefinition}
\end{figure}

After identifying the PoC characteristics founded during the analyzes, it becomes possible to link them with the PoC definition mentioned above, the founded characteristics align with the PoC definition. Figure \ref{fig:pocCharacteristicsDefinition} illustrates the relationship between the PoC characteristics and the established PoC definition.

\section{PoC Characteristics}
\label{sec:characteristics}

Building on the systematic review, eight characteristics emerged as critical for
the PoC creation process (Table~\ref{table:pocCommonCharacteristicsReference}).

\begin{table}[htbp]
\caption{PoC characteristics and supporting references}
\label{table:pocCommonCharacteristicsReference}
\centering
\begin{tabular}{|p{5cm}|p{7cm}|}
\hline
\textbf{Characteristic} & \textbf{References} \\
\hline
Gain confidence / validate design &
  \cite{fairbanks2010just,morris2016infrastructure,burns2018designing,%
        shrivastava2022solutions,baptista2024software} \\
\hline
Quicker experimentation &
  \cite{burns2018designing,ingeno2018software,beningo2022embedded,%
        tune2024architecture} \\
\hline
Solve learning needs &
  \cite{morris2016infrastructure,ingeno2018software,burns2018designing,%
        shrivastava2022solutions,tune2024architecture} \\
\hline
Evaluate risks / discover unknowns / identify gaps &
  \cite{ingeno2018software,burns2018designing,shrivastava2022solutions,%
        tune2024architecture} \\
\hline
Determine feasibility &
  \cite{helgeson2009software,shrivastava2022solutions,tune2024architecture} \\
\hline
PoC should not be a deliverable &
  \cite{ford2017building,burns2018designing,hinchey2018software,%
        shrivastava2022solutions,beningo2022embedded} \\
\hline
Helps to choose / evaluate a proven technology &
  \cite{morris2016infrastructure,ford2017building,burns2018designing,%
        shrivastava2022solutions,brisals2024serverless} \\
\hline
Analyse technical limitations / performance needs &
  \cite{burns2018designing,ingeno2018software} \\
\hline
\end{tabular}
\end{table}

These eight characteristics align with and reinforce our proposed definition.
\textbf{Quicker experimentation} and the \textbf{non-deliverable} nature of PoC
code reinforce its ephemeral character.  \textbf{Solving learning needs} appears
in multiple sources \cite{morris2016infrastructure,ingeno2018software,burns2018designing},
underscoring that a PoC is also a knowledge-generation activity.
\textbf{Determining feasibility} and \textbf{analysing technical limitations}
contribute directly to the decision-making process by providing evidence before
significant investment is committed.

\section{A Structured Framework for PoC Execution}
\label{sec:framework}

\begin{figure}[H]
    \centering
    \fbox{\includegraphics[width=1\linewidth]{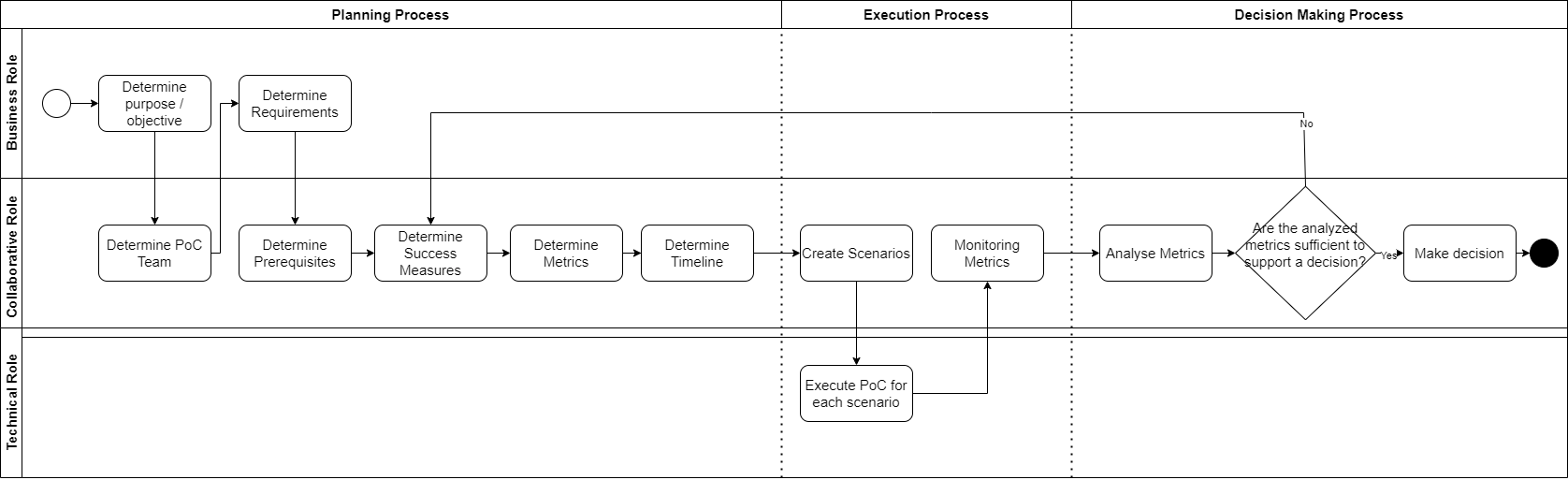}}
    \caption{BPMN PoC}
    \label{fig:BPMNPoc}
\end{figure}

While PoCs should remain lightweight and flexible, we argue that a minimal level of structure is necessary to ensure their architectural value. Drawing on a synthesis of academic work and industrial practice, we outline a simple, repeatable structure that frames PoCs around three interconnected phases: planning, execution, and
decision-making. A detailed BPMN diagram, shown in Figure \ref{fig:BPMNPoc}, provides a comprehensive view of the PoC process.

\subsection{Phase 1: Planning}
\label{sub:planning}

The planning phase serves as the foundation for a PoC.  It makes assumptions explicit by linking the PoC to a concrete decision it is intended to inform.  Seven planning steps were identified:

\begin{itemize}
\item \textbf{Determine purpose/objectives.}  Clearly define the PoC's
hypothesis, whether it concerns a technology, an architectural style, or a
deployment strategy.  Well-defined objectives provide direction and prevent scope
drift.

\item \textbf{Determine personas/stakeholders.}  Identify individuals or groups
directly or indirectly impacted by the PoC, including end-users, project
sponsors, technical teams, compliance officers, and decision-makers.

\item \textbf{Determine requirements.}  Specify the features, functionalities, and constraints the PoC must exercise. Requirements are typically derived from
stakeholder needs and serve as guidelines for validating the initial assumptions.

\item \textbf{Determine prerequisites.}  Document conditions that must be
satisfied before the PoC begins: infrastructure access, data availability,
skilled personnel, budgetary approvals, and tooling licenses.

\item \textbf{Determine success measures.}  Establish clear, measurable criteria
for evaluating outcomes---quantitative metrics (e.g., latency thresholds, cost
savings) and qualitative assessments (e.g., user feedback, stakeholder
satisfaction).

\item \textbf{Determine metrics.}  Specify the concrete data points that
operationalize success measures, for example system response time, user error
rate, or CPU usage.

\item \textbf{Determine timeline.}  Define logical phases rather than fixed
delivery dates, given the inherent unknowns in a PoC, including key milestones,
stopping rules, and resource allocation.
\end{itemize}

\subsection{Phase 2: Implementation and Technical Validation}
\label{sub:execution}

This phase translates planning assumptions into controlled experiments. The first step is to \textbf{create scenarios}: specific use cases that simulate real-world situations and exercise the architectural concerns under investigation, tailored to test core behaviours under varying conditions. The PoC is then \textbf{executed} by deploying the proposed solution in a representative, constrained environment---the goal is to generate empirical data, not to build a production system. Throughout execution, \textbf{metrics are continuously monitored}: tracking the KPIs defined in the planning phase enables real-time identification of issues and timely adjustments before the decision-making phase begins.

\subsection{Phase 3: Decision-Making}
\label{sub:decision}

The decision-making phase translates experimental results into architectural
action. The collected data are first \textbf{analysed} against the predefined
success criteria to identify trends, trade-offs, and limitations. Based on this
analysis, a concrete \textbf{decision} is made---to adopt, reject, or revise
the architectural approach under evaluation.

At this point, PoC implementation artifacts may be discarded \emph{from production codebase}; architectural knowledge, however, must be preserved as technical evidence.

\section{Framework Document Structure}
\label{sec:framework-structure}

At a practical level, the framework is materialized as a structured document organized into four sections.
Section~(1), \textit{General information}, captures the PoC's purpose, the team involved, and its key characteristics, establishing the context and motivation for the experiment.
Section~(2), \textit{Requirements, prerequisites, success measures, and metrics}, specifies what must be true before the PoC begins, what it must demonstrate, and how outcomes will be measured.
Section~(3), \textit{Timeline and PoC conclusion}, defines the logical phases of execution and records the final decision together with its architectural rationale.
Section~(4), \textit{Scenario description and evaluation}, documents the concrete test scenarios and the empirical results collected during execution.

Together, these four sections transform a PoC from an informal experiment into a traceable architectural artifact.
A reusable template implementing this structure is available online.\footnote{\url{https://docs.google.com/spreadsheets/d/1o0T8DBjjQbyAQgNdqhHnRSUwiKKMpOt2LQDejM_s248}}
This template is applied in the following section to evaluate a real industrial case study in the banking domain and to conduct a reverse-engineering exercise on a publicly available performance benchmark.

\section{Framework Evaluation: Industrial Case Studies}
\label{sec:casestudies}

We evaluated the framework through two case studies.  The first applies the
framework to an industrial PoC in a regulated banking domain; the second conducts
a reverse-engineering analysis of a publicly available performance benchmark.

\subsection{Case Study 1: Database Versioning in a Banking Environment}
\label{sub:banking}

\subsubsection{Context.}
Database versioning management is a critical requirement for financial
institutions driven by operational needs and strict regulatory demands.  This
case study evaluates two widely-used database versioning tools---Flyway and
Liquibase---within a Java-based banking application using documentation from the
bank's internal Confluence pages as the primary information source.

\subsubsection{Applying the Framework.}

\paragraph{Planning.}
The PoC objective was to determine which tool better supports schema versioning, rollback mechanisms, and compliance with banking regulations.  Four stakeholder
groups were identified: Database Administrators (DBAs), Compliance Officers, Developers, and IT Managers.  Functional requirements included version control for
schema and data changes, rollback mechanisms, and compatibility with Oracle and PostgreSQL.  Non-functional requirements focused on minimizing migration downtime,
performance under high transaction volumes, and CI/CD integration.  Prerequisites included isolated database instances, prepared migration scripts, configured
CI/CD pipeline, network access, administrator privileges, and a valid Docker license.

Success measures were: migration time, audit-log completeness and clarity, and
rollback effectiveness.  Corresponding metrics were measured in seconds/minutes,
assessed against compliance standards, and benchmarked against prior rollback
scenarios.

The timeline was divided into four logical stages: environment setup, Flyway
execution, Liquibase execution, and comparative analysis.

\paragraph{Implementation and technical validation.}
Eight DDL and DML scripts were designed to test structural changes, parallel
development conflicts, rollback, complex schema changes, data migration, and
production-like downtime minimization.  Both tools were executed against these
scenarios, and all metrics were recorded in Confluence for auditability.

\paragraph{Decision-making.}
Liquibase demonstrated superior auditability and more robust rollback features.
However, the development team preferred Flyway due to its simpler integration
with SQL syntax---an important usability consideration aligned with existing
practices.  Critically, when raw SQL migrations were used (supported by both
tools), Liquibase's rollback advantage disappeared, as rollback operations
required manual intervention.  The final recommendation was \textbf{Flyway},
which provided a faster, more intuitive solution that met compliance requirements
in a more straightforward manner.

\subsubsection{Framework Contribution.}
To assess the framework's value, we compared the artefacts produced by the
original unstructured PoC (as documented in the bank's internal Confluence pages)
against those produced after applying the framework, across three dimensions:
\textbf{completeness of requirements}, \textbf{clarity of success criteria},
and \textbf{decision traceability}.  In all three dimensions, the structured
artefacts showed measurable improvement: role distinctions among stakeholders
were absent in the original documentation and explicitly defined after applying
the framework; functional and non-functional requirements, which were conflated
in the Confluence pages, were clearly separated in the structured version; and
success measures that were implicit in the original were made explicit and tied
directly to compliance standards.  The structured evaluation facilitated a
collaborative, evidence-based decision that would not have been achievable with
the original unstructured documentation alone.

\subsection{Case Study 2: Performance Benchmarking Java vs.\ Golang on Kubernetes}
\label{sub:java_golang}

\subsubsection{Context.}
This case study applies the framework through \emph{reverse engineering} to a
publicly available YouTube tutorial that benchmarks Java~(Quarkus) and
Golang~(Fiber) applications on Kubernetes \cite{anton_putra_performance_benckmark_2024}.
The grey-literature quality assessment (authority, purpose, methodology,
relevance) following \cite{GAROUSI2019101} rated the source 17.5/20, confirming
its credibility.

\subsubsection{Applying the Framework.}

\paragraph{Planning gaps identified.}
The original study did not define stakeholders, requirements, prerequisites, or a
timeline.  Applying the framework allowed us to infer three stakeholder groups
(DevOps Engineers, Cloud Architects, Developers) and to specify functional
requirements (CPU/memory monitoring, latency measurement, startup time
evaluation) and non-functional requirements (scalability, cost efficiency).
Prerequisites included Prometheus, Grafana, AWS S3, PostgreSQL, a Kubernetes
cluster, and a GitHub repository.

The framework also surfaced three additional success measures not present in the
original study: Latency Performance, Resource Efficiency, and Scalability; and
three additional metrics: image size, file operation latency, and database
operation latency.

\paragraph{Scenario expansion.}
The framework led to three structured test scenarios: (i)~REST API response time
under varying load; (ii)~file handling and database interaction with S3 and
PostgreSQL; and (iii)~application startup time and scalability under load
increases.  Scenario~(iii) was new, emerging from the image-size metric.

\paragraph{Decision-making.}
Golang~(Fiber) demonstrated lower latency (p99), faster startup time, and a
significantly smaller Docker image, making it preferable for cloud-native
deployments requiring rapid scaling.  Java~(Quarkus) is viable for organizations
already invested in the Java ecosystem.

\subsubsection{Framework Contribution.}
The reverse-engineering exercise confirms the framework's ability to structure
PoCs that were originally informal, to uncover missing requirements and metrics,
and to produce decisions with explicit architectural rationale.

\section{PoCs as Architectural Decision Instruments}
\label{sec:architectural}

\subsection{Reframing PoCs}

We propose a reframing of PoCs in software architecture.  Rather than treating
PoCs as informal or peripheral activities, we conceptualize them as
\emph{short-term, scoped experiments explicitly designed to validate
architectural hypotheses}.  From this perspective, a PoC is primarily a
decision-oriented artifact: its main outcome is not executable code, but
\emph{evidence} that supports or refutes an architectural choice.

This reframing highlights four essential characteristics:
(i)~\textbf{Decision-driven}---PoCs are initiated to inform a concrete
architectural decision;
(ii)~\textbf{Risk-driven}: they focus on the most uncertain or high-risk
aspects of an architecture where analytical reasoning alone is insufficient;
(iii)~\textbf{Ephemeral}: they are not intended to become production assets;
(iv)~\textbf{Architecturally scoped}: they target specific architectural
concerns such as performance, scalability, security, interoperability,
compliance, or operational constraints.

\subsection{Integration with Architectural Knowledge Artifacts}

Proofs of Concept naturally align with the intent of Architecture Decision
Records~\cite{fowler2026adr}.  While ADRs document significant architectural
decisions and their rationale, PoCs often generate the empirical evidence on
which those decisions are implicitly based.  In the Advice Process model
\cite{fowler2023scaling}, ADRs serve not only as records but as thinking tools
that elicit and consolidate expertise before a decision is finalised.  PoCs are
the experiment that generates the evidence that feeds this process---yet in
practice the relationship between PoCs and ADRs is rarely made explicit: PoCs
conclude, architectural choices are made, but the assumptions tested, the metrics
observed, and the trade-offs identified during experimentation are frequently
decoupled from the architectural record.

\subsection{The Undocumented Architectural Experiment Anti-Pattern}
We introduce the \textbf{Undocumented Architectural Experiment}  anti-pattern, defined as the habitual separation between Proofs of Concept (PoCs) and Architecture Decision Records (ADRs).

ADRs are meant to capture the rationale behind significant architectural choices, yet PoCs are executed precisely to generate the
evidence that determines those choices.  When PoCs conclude without producing,
refining, or invalidating an ADR, architectural decisions are effectively
\emph{made off the record}.  The result is a form of ``ghost architecture,''
where high-impact choices are justified by experiments that leave no durable
trace of their assumptions, metrics, or trade-offs.

This anti-pattern produces several systemic consequences. Most critically, it
leads to the progressive \textbf{loss of architectural knowledge}, as
experimental insights remain undocumented and cannot be reused. It also results
in \textbf{non-reproducible outcomes}, where experiments cannot be repeated or
verified due to missing contextual information. \textbf{Decision traceability}
becomes weak, making it difficult to justify architectural choices
retrospectively, particularly in regulated or safety-critical environments.
Over time, these effects reduce \textbf{organizational learning}, as knowledge
generated through experimentation fails to accumulate into durable architectural
assets.

The presence of this anti-pattern can often be detected through recurring
symptoms in practice. Teams may acknowledge that an experiment was performed but
lack a clear description of how it was conducted. Results may exist without
explicit statements of the assumptions or evaluation criteria that produced
them. In more critical cases, architectural decisions cannot be justified
retrospectively because the experimental evidence that informed them was never
formally captured. These symptoms indicate that experimentation is occurring
\emph{outside} the architectural knowledge lifecycle rather than contributing
to it.

A PoC that neither produces an Architecture Decision Record nor invalidates a
candidate ADR should therefore be considered \textbf{architecturally
incomplete}, regardless of how technically successful the experiment may appear.
Framed in this way, PoCs become accountable architectural acts---\emph{empirical
gates} in the architectural decision lifecycle---determining whether a candidate
decision is promoted, revised, or rejected.

\subsection{Open Challenges}

Treating PoCs as first-class architectural decision instruments raises several
open challenges. A first tension concerns \textbf{standardization vs.\ flexibility}.  Structure is necessary to ensure decision quality, but excessive formalization risks turning PoCs into heavyweight processes that undermine their exploratory nature. The framework must therefore remain lightweight and adaptable to the unpredictability inherent in experimentation.

A second challenge is \textbf{tooling integration}.  There is an opportunity to better leverage DevOps and observability platforms to support PoC-driven decision-making and continuous architectural evaluation. Tighter tooling integration would lower the cost of structured experimentation and improve the traceability of results.

Third, \textbf{PoC practices must adapt to contemporary contexts} such as
cloud-native systems, AI/ML architectures, and regulated domains, each of which presents unique constraints that require tailored experimentation strategies beyond what current frameworks address.

Finally, the relative absence of PoCs from software architecture research
points to a \textbf{broader research gap} between architectural theory and
practice. Empirical studies--controlled experiments, longitudinal field
studies, and structured practitioner surveys--are needed to validate, refine,
and generalize the findings presented here.

\section{Discussion}
\label{sec:discussion}

\subsection{Answers to Research Questions}

\paragraph{RQ1: What is the definition of PoC in software engineering?}
This research provides a refined definition that addresses the conceptual
ambiguity found in the literature.  A PoC is a \emph{targeted, time-constrained
experiment} aimed at reducing technical risks, supporting informed
decision-making, and validating specific hypotheses before committing substantial
resources to full-scale development.  It is distinct from prototypes (user
experience focus) and MVPs (market viability focus).

\paragraph{RQ2: What characteristics and processes contribute to effective PoC
planning, execution, and evaluation?}
Eight key characteristics emerged (Table~\ref{table:pocCommonCharacteristicsReference}).
The PoC process is structured around three phases: planning, execution, and
decision-making.  These phases interact continuously, allowing the PoC to adapt
dynamically based on emerging insights.  This research recommends implementing
these phases in small, incremental steps to provide flexibility and foster
continuous learning.

\paragraph{RQ3: What are the benefits and challenges of a structured PoC
framework?}
The framework enhances consistency and clarity by providing step-by-step guidance
for practitioners, aiding in resource management, and supporting decision-making
through clear evaluation criteria.  Challenges include the difficulty of
generalizing the framework to all project types and the risk of scope creep.
The framework's feedback loops and clearly defined objectives help address these
challenges.

\subsection{Implications for Practice}

The framework provides software engineering professionals with a structured
approach to planning and executing PoCs, ensuring that stakeholder engagement,
success metrics, and scenario planning are adequately addressed.  This
contributes to better resource allocation, improved risk management, and higher
project success rates.  By clearly defining each stage of the PoC process, the
framework helps practitioners avoid common pitfalls such as scope creep,
misaligned objectives, and inadequate testing.  Crucially, by linking PoC
outcomes to architectural decision records, practitioners can build a durable
body of architectural knowledge rather than allowing critical decisions to remain
undocumented.

\subsection{Limitations}

First, as in any qualitative study, potential biases in data collection and
interpretation may have influenced the findings.  Second, the literature review,
while systematic, may not have captured all emerging or unpublished work.
Third, PoC implementation may require significant time, expertise, and budget,
which may limit the framework's applicability in resource-constrained teams.
Fourth, validation of the framework is based on two case studies; additional
empirical validation---controlled experiments, structured interviews, and
longitudinal studies---is required to generalize the findings.

\subsection{Future Work}

Future research should focus on further validating the framework in live software
development projects and comparing PoC success rates with and without the
framework.  The framework could be expanded to include emerging technologies such
as AI/ML systems and blockchain, which present unique challenges for PoC
implementation.  Longitudinal studies across different industries would help
assess the framework's applicability and identify necessary adaptations.

\section{Conclusion}
\label{sec:conclusion}

This paper addresses the ambiguity surrounding the definition and implementation
of Proof of Concept in software engineering.  We proposed a refined definition
that distinguishes PoCs from prototypes, pilots, and MVPs; identified eight key
characteristics grounded in systematic literature analysis; and introduced a
lightweight, three-phase framework (planning, execution, decision-making) that
guides practitioners through the entire PoC lifecycle.

Two industrial case studies demonstrated that the framework surfaces hidden
requirements, clarifies stakeholder roles, aligns metrics with success criteria,
and structures decision-making---improvements that were absent or implicit in the
original PoC executions.

We further argued that PoCs should be treated as \emph{first-class
architectural decision instruments}: short-lived, risk-driven experiments whose
primary output is evidence rather than code.  We introduced the
\emph{Undocumented Architectural Experiment} anti-pattern to name the common
failure mode in which PoCs influence high-impact architectural decisions without
leaving durable architectural knowledge.

Reframing PoCs this way---making their assumptions, evidence, and decision points
explicit---is a necessary step toward more empirical, accountable, and
continuously evolving software architectures.  If PoCs continue to play a
decisive role in shaping software systems, we cannot afford to let their
influence remain undocumented and largely invisible.

\paragraph{Data availability.}
Case Study~1 is derived from internal artefacts of a regulated financial
institution; the underlying data cannot be shared publicly for confidentiality
reasons.  Case Study~2 is based on a publicly available YouTube tutorial
\cite{anton_putra_performance_benckmark_2024}.  The PoC framework template
applied in both case studies is available at the URL provided in
Section~\ref{sec:framework-structure}.

\bibliographystyle{splncs04}
\bibliography{references}

%
% ...

\end{document}